\documentclass[sigconf]{acmart}
\usepackage[utf8]{inputenc}
\usepackage{graphicx}
\usepackage{subcaption}
\usepackage{CJKutf8}
\usepackage{enumitem}
\usepackage{hyperref}
\usepackage{booktabs}
\usepackage{multirow}
\usepackage{placeins}
\AtBeginDocument{%
  }

\copyrightyear{2026}
\acmYear{2026}
\setcopyright{cc}
\setcctype{by}
\acmConference[CHIWORK '26]{Adjunct Proceedings of the 5th Annual Symposium on Human-Computer Interaction for Work}{June 22--25, 2026}{Linz, Austria}
\acmBooktitle{Adjunct Proceedings of the 5th Annual Symposium on Human-Computer Interaction for Work (CHIWORK '26), June 22--25, 2026, Linz, Austria}
\acmDOI{10.1145/3805029.3818282}
\acmISBN{979-8-4007-2598-2/2026/06}
\begin{document}

%%
%% The "title" command has an optional parameter,
%% allowing the author to define a "short title" to be used in page headers.
%\title{Effortless Navigation for Individuals with Visual Impairments: A Real-time AI-Driven Mobile System (NaviGPT)}
%\title{Human-AI Collaboration in Intelligent Augmented Environments: Theory, Practice, and Challenges}
\title[Ethics and Responsibility in AI-Assisted Interviewing]{Ethics and Social Responsibility in AI-Assisted Interviewing: \\An LLM-in-the-Loop Study of AI-Generated Follow-Up Questions}
%\title{Enhancing Human Capabilities through AI-Powered Augmentation}%: A Framework for Designing and Evaluating LLM-based Collaborative Systems}
%%
%% The "author" command and its associated commands are used to define
%% the authors and their affiliations.
%% Of note is the shared affiliation of the first two authors, and the
%% "authornote" and "authornotemark" commands
%% used to denote shared contribution to the research.
\author{He Zhang}
\orcid{0000-0002-8169-1653}
\affiliation{%
  \institution{College of Information Sciences and Technology, The Pennsylvania State University}
  \city{University Park}
  %\state{Pennsylvania}
  \state{PA}
  \country{USA}
  \postcode{16802}
}
\email{hpz5211@psu.edu}

\author{Yueyan Liu}
\orcid{}
\affiliation{%
  \institution{The Future Laboratory, Tsinghua University}
  \city{Beijing}
  \country{China}
  \postcode{100084}
}
\email{yueyan_liu@mail.tsinghua.edu.cn}

\author{Xin Guan}
\orcid{}
\affiliation{%
  \institution{Teachers College, Columbia University}
  \city{New York}
  \state{NY}
  \country{USA}
  \postcode{10027}
}
\email{xg2413@tc.columbia.edu}

\author{Jie Cai}
\authornote{Corresponding author}
\orcid{0000-0002-0582-555X}
\affiliation{%
  %\institution{Pervasive Human Computer Interaction Laboratory, Department of Computer Science and Technology, Tsinghua University}
  \institution{Department of Computer Science and Technology, Tsinghua University}
  \city{Beijing}
  \country{China}
  \postcode{100084}}
\email{jie-cai@mail.tsinghua.edu.cn}

%\author{ChanMin Kim}
%\email{cmk604@psu.edu}
%\orcid{0000-0001-9383-8846}
%\affiliation{%
%  \institution{College of Education, Penn State University}
%  \city{University Park}
%  \state{Pennsylvania}
%  \country{USA}
%  \postcode{16802}
%}

\author{John M. Carroll}
\authornotemark[1]
\orcid{0000-0001-5189-337X}
\affiliation{%
  \institution{College of Information Sciences and Technology, The Pennsylvania State University}
%   \streetaddress{1 Th{\o}rv{\"a}ld Circle}
  \city{University Park}
  \state{PA}
  \country{USA}
  \postcode{16802}}
\email{jmc56@psu.edu}

%%
%% By default, the full list of authors will be used in the page
%% headers. Often, this list is too long, and will overlap
%% other information printed in the page headers. This command allows
%% the author to define a more concise list
%% of authors' names for this purpose.
\renewcommand{\shortauthors}{Zhang et al.}

%%
%% The abstract is a short summary of the work to be presented in the
%% article.
\begin{abstract}
Semi-structured interviews rely on timely, context-sensitive follow-up questions, yet interviewers' cognitive load and limited domain familiarity can constrain probing depth. We report findings from an LLM-in-the-loop Wizard-of-Oz (WoZ) study that simulates an AI follow-up assistant in live interviewing while preserving human oversight. In our setup, a co-interviewer selectively relayed and could edit AI-generated follow-up questions (AGQs) produced in real time by GPT-4o, enabling a realistic approximation of deployment without fully automating the interaction. Across 17 interviewers with varied qualitative-method expertise, participants raised five interlocking concerns: (1) harmful or discriminatory language and unpredictable interaction harms, (2) undermining interviewees' sense of respect through divided attention and missing nonverbal cues, (3) technology-based participation inequality, (4) unclear responsibility when harms occur, and (5) privacy, disclosure, and compliance risks when AI listens, records, or transcribes sensitive content. We translate these concerns into design and governance implications for safer, more respectful, and more accountable AI-assisted interviewing.

%Semi-structured interviews highly rely on the quality of follow-up questions, yet interviewers' knowledge and skills may limit their depth and potentially affect outcomes. While many studies have shown the usefulness of large language models (LLMs) for qualitative analysis, their possibility in the data collection process remains underexplored. We adopt an AI-driven ``Wizard-of-Oz'' setup to investigate how real-time LLM support in generating follow-up questions shapes semi-structured interviews. Through a study with 17 participants, we examine the value of LLM-generated follow-up questions, the evolving division of roles, relationships, collaborative behaviors, and responsibilities between interviewers and AI. Our findings (1) provide empirical evidence of the strengths and limitations of AI-generated follow-up questions (AGQs); (2) introduce a Human-AI collaboration framework in this interview context; and (3) propose human-centered design guidelines for AI-assisted interviewing. We position LLMs as complements, not replacements, to human judgment, and highlight pathways for integrating AI into qualitative data collection.
\end{abstract}

%%
%% The code below is generated by the tool at http://dl.acm.org/ccs.cfm.
%% Please copy and paste the code instead of the example below.
%%

%\ccsdesc[500]{General and reference~Cross-computing tools and techniques}
%%
%% Keywords. The author(s) should pick words that accurately describe
%% the work being presented. Separate the keywords with commas.

%\received{20 February 2007}
%\received[revised]{12 March 2009}
%\received[accepted]{5 June 2009}

\begin{teaserfigure}
    \centering
    \includegraphics[width=0.6\linewidth]{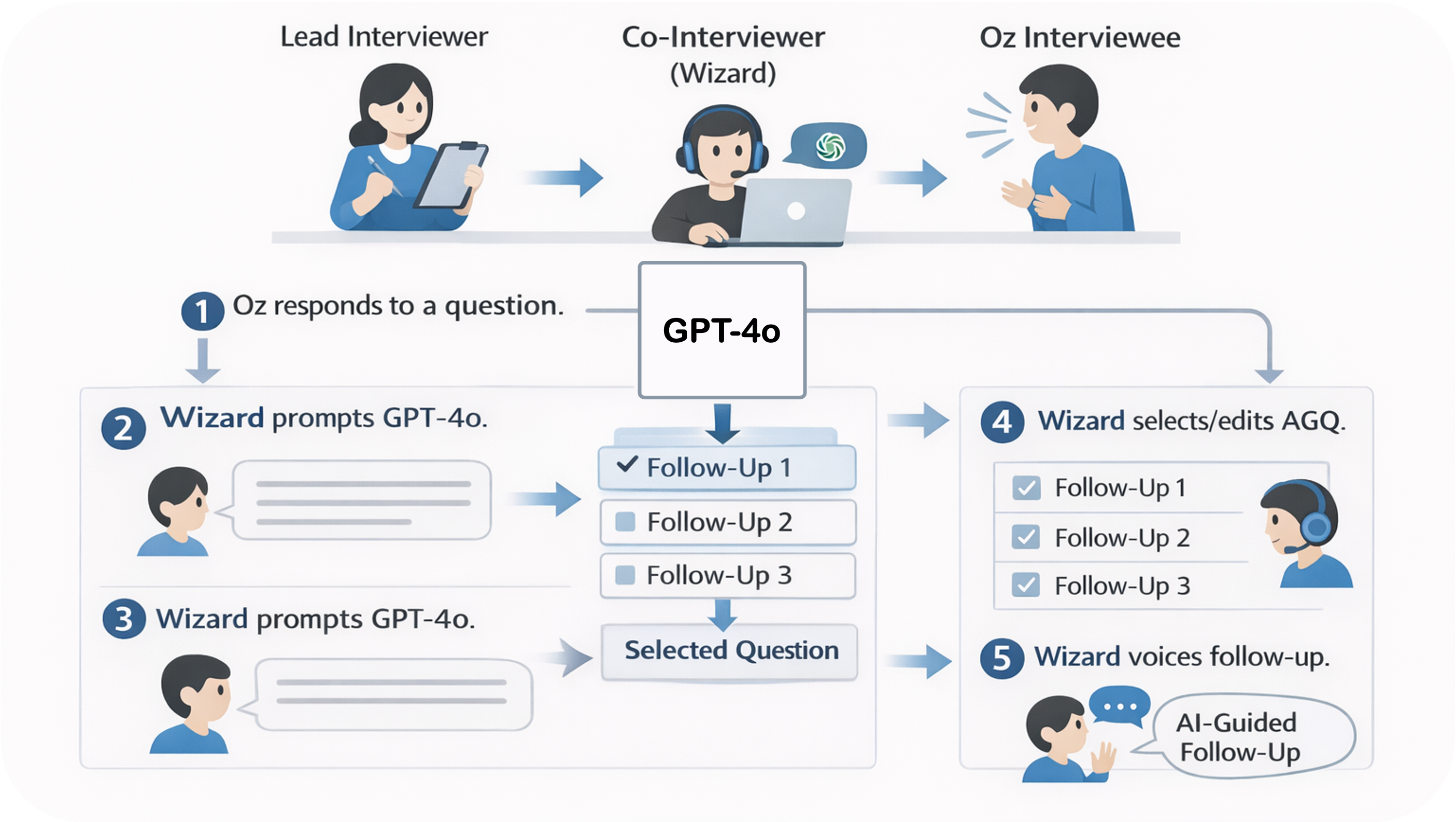}
    \caption{LLM-in-the-loop Wizard-of-Oz study design for AI-assisted qualitative interviewing (study schematic). The lead interviewer conducts a semi-structured interview with an ``Oz'' interviewee. The Wizard requests 1-3 candidate follow-ups from GPT-4o based on Oz's response, then selectively voices/edits a follow-up. This simulates an AI follow-up assistant with human oversight.}
    % 
    %\Description[]{}
    \label{fig:teaser}
\end{teaserfigure}

%\begin{teaserfigure}
%    \centering
%    \includegraphics[width=0.9\linewidth]{graph/teaserfigure.png}
%    \caption{Schematic of the AI-driven Wizard-of-Oz method used in this study. The left panel shows the role configuration in the interview setup, where the Human co-interviewer's actual questions are generated by the AI. The right panel illustrates the interaction flow among the different parties during the interview.}
%    % 
%    %\Description[]{}
%    \label{fig:teaser}
%\end{teaserfigure}
%%

%\begin{teaserfigure}
%\centering
%\includegraphics[width=0.9\textwidth]{graph/workflow.png}
%\caption{Workflow of NaviGPT. }
%\Description{}
%\end{teaserfigure}
%% This command processes the author and affiliation and title
%% information and builds the first part of the formatted document.
\maketitle
\vspace{-0.3cm}
\section{Introduction}

Semi-structured interview is the one of the most common format of data collection and one of the cornerstone of qualitative research, relying on interviewers' ability to produce timely, context-sensitive follow-up questions that probe emerging themes while sustaining rapport and trust~\cite{roulston2010,king2018interviews,jamshed2014qualitative}. Generating such follow-ups in real time is cognitively demanding and uneven across levels of methodological expertise, particularly under time pressure or when addressing complex and sensitive topics~\cite{adams2015,flick2006,marshall2014}. Recent advances in large language models (LLMs) have therefore motivated growing interest in AI-assisted interviewing tools that suggest candidate follow-up questions during live interaction~\cite{10.1145/3382507.3418839,10.1145/3686215.3688377,liu2025aiinterviews,spangher-etal-2025-newsinterview,zhang2025harnessingpoweraiqualitative}.

While prior research has largely examined the utility and perceived helpfulness of AI-generated follow-up questions (AGQs), less attention has been paid to the ethical and social responsibility implications of integrating AI assistance into the moment-by-moment dynamics of interviews. Qualitative interviews are not merely information-extraction procedures; they are social encounters shaped by trust, power relations, respect, and accountability~\cite{kvale2006dominance,karnieli2009power}. Introducing AI systems into these interactions, directly or indirectly, may alter how responsibility, agency, and ethical obligations are understood by interviewers and interviewees~\cite{10.1145/3581641.3584051,liu2023speech}.

This work investigates these issues through an LLM-in-the-loop Wizard-of-Oz (WoZ) study that simulates an AI follow-up assistant with human oversight. A co-interviewer selectively reviews, edits, and voices candidate follow-up questions generated by GPT-4o in real time, approximating realistic deployment while avoiding fully autonomous AI intervention. The operating procedure is shown in Fig.~\ref{fig:teaser}. Drawing on interviewers' reflections before and after AI involvement was disclosed, we identify five interrelated ethical concerns: language harms and unpredictable interaction risks; erosion of interviewees' sense of respect and rapport; technology-based participation inequality; ambiguity in responsibility attribution when errors occur; and privacy, disclosure, and compliance risks. Together, these findings foreground AI-assisted interviewing as a socio-technical practice that demands careful ethical design and governance.

This study is guided by two research questions:
\begin{itemize}
    \item[RQ1.] How do interviewers perceive the ethical and social responsibility implications of AI-generated follow-up questions in a simulated semi-structured interviewing context?

    \item[RQ2.] What design and governance considerations do interviewers' reflections suggest for future AI-assisted interviewing tools?
\end{itemize}

\vspace{-0.4cm}
\section{Method}
\vspace{-0.1cm}
\subsection{Participants}
%We recruited 17 participants with prior experience in conducting or supporting semi-structured interviews. Participants were recruited through the distribution of study flyers on social media platforms and through peer referrals within the human–computer interaction and social science research communities. This recruitment strategy enabled us to include interviewers with varied levels of qualitative research experience, ranging from novices with limited formal training to experienced researchers who regularly conduct interviews as part of their scholarly work (Novice (10) / Intermediate (5) / Advanced (2)).

%Participants came from diverse disciplinary backgrounds, including human–computer interaction, information sciences, and social sciences. All study sessions were conducted remotely via video conferencing. Participation was voluntary, and all participants provided informed consent prior to the study. %Table~\ref{tab:participants} summarizes sample characteristics.

We recruited 17 participants who had prior experience conducting or supporting qualitative research. Participants were recruited through study flyers distributed on social media and peer referrals within HCI and social science research communities. The final sample included 10 women and 7 men, aged 21--35. Participants came from fields including HCI, education, psychology, social work, applied linguistics, human geography, management science, usable privacy and security, and humanities/social sciences.

We categorized participants' qualitative-method experience into three levels. Novice participants had participated in qualitative data collection, especially interviews. Intermediate participants had published one to three peer-reviewed papers or reports based on qualitative methods. Advanced participants had published more than three peer-reviewed qualitative-method papers. Based on this definition, the sample included 10 novice, 5 intermediate, and 2 advanced participants.

We recognize that this distribution overrepresents novice and intermediate researchers. We therefore interpret the findings primarily as reflecting how researchers with varied but often early-stage qualitative-method experience reason about AI-assisted interviewing, rather than as a comprehensive account of expert qualitative researchers' ethical practice.

%\vspace{-0.1cm}
%\begin{table}[htbp]
%\small
%\centering
%\caption{Participant summary (replace with exact counts).}
%\vspace{-0.2cm}
%\label{tab:participants}
%\begin{tabular}{@{}p{0.33\linewidth}p{0.58\linewidth}@{}}
%\toprule
%\textbf{Characteristic} & \textbf{Summary} \\
%\midrule
%Sample size & 17 interviewers \\
%Expertise levels & Novice (10) / Intermediate (5) / Advanced (experience-based) (2) \\
%Prior interviewing & Varied (coursework, research projects, applied settings) \\
%Interview modality & Remote video conferencing \\
%\bottomrule
%\end{tabular}
%\end{table}

\vspace{-0.2cm}
%\subsection{Procedure}
%Each session involved three roles: (1) a recruited participant acting as the lead interviewer, (2) a researcher acting as the interviewee (``Oz'') to standardize interaction content, and (3) a researcher acting as the co-interviewer (``Wizard'') who selectively introduced AGQs generated by GPT-4o in real time. We varied the timing pattern of Wizard follow-ups (e.g., occasional insertion; periodic insertion; clustered insertion) to approximate realistic co-interviewing rhythms.

%Crucially, participants completed a two-stage reflection: they first evaluated the co-interviewer's follow-ups \emph{without} being told those follow-ups originated from an LLM; we then disclosed LLM involvement and conducted a second reflection focused on disclosure, trust, ethics, and responsibility.
\subsection{Procedure}
Before starting the study, we asked each participant to provide an interview topic they wished to discuss, along with a general direction for several interview questions, and we fed this information into the LLM as initial context. Each session then involved three roles: (1) a recruited participant acting as the lead interviewer, (2) a researcher acting as the interviewee (``Oz'') to standardize interaction content, and (3) a researcher acting as the co-interviewer (``Wizard''). During each session, the Wizard entered Oz's most recent response into GPT-4o and requested one to three candidate AI-generated follow-up questions (AGQs), using the prompt: ``Please generate follow-up questions based on the participant's responses: *transcription of the participant's response content*.'' The Wizard then selected whether and when to voice one candidate question. We varied the timing pattern of Wizard follow-ups (e.g., occasional insertion; periodic insertion; clustered insertion) to approximate realistic co-interviewing rhythms. The Wizard was instructed not to introduce new substantive content beyond the AI-generated candidate question.

We varied the timing pattern of Wizard follow-ups across three interaction patterns: occasional single follow-ups, periodic follow-ups after one or several participant-led questions, and clustered follow-ups in which two or more AGQs were asked consecutively.  Crucially, participants completed a two-stage reflection: they first evaluated the co-interviewer's follow-ups without being told those follow-ups originated from an LLM; we then disclosed LLM involvement and conducted a second reflection focused on disclosure, trust, ethics, and responsibility.

Lastly, because the interviewee (``Oz'') was played by a researcher in this study, the conversational content entered into GPT-4o for AGQ generation consisted of Oz's responses within the simulated interview rather than personal disclosures from recruited participants. Thus, although participants interacted with AI-generated follow-up questions, we did not submit recruited participants' personal interview responses or sensitive personal data to the LLM for AGQ generation. All participants provided informed consent before taking part in the study.

\vspace{-0.2cm}
\subsection{Data and Analysis}
With participants' consent, all sessions were audio-recorded and transcribed verbatim. Transcripts and associated materials were anonymized prior to analysis. We conducted an inductive thematic analysis using a consensus-oriented coding process.

The dataset contains participants' reflections on a range of experiences with AI-assisted interviewing. In this work, we deliberately center our analysis on themes related to ethics and the allocation of social responsibility, including concerns about language harm, respect and rapport, participation inequality, responsibility attribution, and data privacy. More specifically, during open coding, we marked segments in which participants discussed potential harms, trust, respect, responsibility, privacy, disclosure, control, and appropriate use of AGQs. We then met to compare codes, merge overlapping labels, and develop a shared codebook. Next, we applied the refined codes across the full dataset, writing analytic memos to capture recurring patterns and tensions. The five themes reported in this paper were developed through iterative clustering of codes that repeatedly appeared across participants and were directly connected to ethical or social responsibility concerns. For example, comments about offensive wording, discriminatory phrasing, and unpredictable model behavior were grouped under language risks and interaction harms; comments about divided attention, nonverbal cues, and interviewee comfort were grouped under respect and rapport; comments about digital literacy and unequal access were grouped under participation inequality; comments about fault, blame, and human oversight were grouped under responsibility ambiguity; and comments about recording, transcription, disclosure, and third-party AI services were grouped under privacy and compliance. Disagreements were resolved through discussion until consensus was reached.

\vspace{-0.3cm}
\section{Findings: Interviewers’ Perceived Ethical Tensions in AI-Assisted Interviewing}
Participants articulated five ethical concerns when imagining AGQs as a practical interviewing aid. 

\vspace{-0.3cm}
\subsection{Language Risks and Interaction Harms}

Participants expressed concern that AI systems may generate abnormal, offensive, or aggressive responses during interview interactions. They emphasized unpredictability under subtle adversarial cues. For example, one participant emphasized that, unlike humans who may ``\textit{perform poorly}'' but  ``\textit{not talk nonsense}'', AI systems could produce ``\textit{very bizarre remarks}'' when triggered in unpredictable ways (P15).

Some also worry about discriminatory or culturally inappropriate language that would be generated by AI and difficult to repair in real time. As one participant noted,
\textit{``It would be very troublesome if AI said to an immigrant, `Your accent is really unpleasant'. I wouldn't even know how to smooth things over''} (P16).
In addition to the linguistic appropriateness, participants were also worried about boundary violations and conflict escalation. Aggressive responses were seen as potentially ``\textit{violating the boundaries of the interviewee},'' particularly in sensitive interview contexts (P1). These concerns show participants' anxiety about AI's unpredictability and limited capability to repair social and relational harm in the interview process.

Conversely, a minority argued that when following established language guidelines, AI often poses appropriate questions and can make fewer mistakes than humans, especially when a human can intervene (P13, P17). This contrast highlighted that the core tension was reliability under real-time pressure.
\vspace{-0.2cm}
\subsection{Undermining the Interviewee's Sense of Respect}
Several participants highlighted that AI follow-up assistants may introduce subtle forms of disrespect and discomfort for interviewees. Monitoring AI outputs can divide the interviewer's focus and be interpreted as disengagement (e.g., ``eyes wandering in another direction'') (P7). Participants also linked respect to nonverbal cues (eye contact, facial expressions) that AI lacks:
\textit{``AI does not have the real expressions of a normal human being. One can perceive certain atmospheres and emotions from a human, but AI can never do that''} (P7).
From an interviewee's perspective, AI-driven interviewing can feel rigid and procedural. Participants explained that once a time limit is reached, the system ``automatically moves on to the next question'' (P7). This procedural rigidity was regarded as harmful to creating a respectful experience for the interviewees.

Such concern was amplified for vulnerable groups as interviewees, where rapport and trust are difficult to establish and easy to damage. In this context, P10 emphasized the challenge of making initial contact with those groups and further highlighted the need to avoid ``offending them'' or ``crossing their bottom lines throughout the process'' (P10).  

Furthermore, for some participants, these interactional issues raised broader ethical questions about the legitimacy of AI-to-human questioning itself. For instance, P7 questioned the ethical appropriateness of AI-to-human questioning:
\textit{``AI-to-Human is very impolite and disrespectful, it should be either human-to-human or AI-to-AI. It's just the basic respect''} (P7).
\vspace{-0.2cm}
\subsection{Technology-Based Participation Inequality}
Participants also worried that AI assistants may introduce implicit forms of technical discrimination, particularly by privileging those who know how to interact with such systems effectively. For example, one participant drew analogies to AI customer service that requires knowing ``\textit{the right keywords}'':
\textit{``If you don't know the keywords, you won't get the desired service. For those who don't understand technology, this is equivalent to a kind of knowledge-based discrimination.''} (P7)
To mitigate such risks, some adopted more conservative approaches and limited reliance on AI due to unequal technological ability. P16, for example, chose not to ``\textit{use GPT much}'' and instead relied on ``\textit{more traditional methods}'' in order to accommodate differences in technological ability across interviewees (P16). This strategy suggests an effort to reduce technology-based participation inequality by accommodating different levels of digital literacy.

\vspace{-0.3cm}
\subsection{Unclear Responsible Subject}
When the system generates inappropriate or offensive content, questions of responsibility become unavoidable. Participants noted that the social meaning of errors differs depending on whether the AI or a human takes the responsibility. P14 emphasized this difference by explaining an example in the interviewing setting: 
\textit{``I think if it's a machine that says something wrong, I have more reason not to pay attention to it... But if it's a human, it will make the interviewee feel that we aren't fully prepared, which is irresponsible.''} (P14)
This subtle difference in responsibility attribution was further seen as contingent on how the AI is framed and configured. For instance, P14 deliberately differentiates between cases involving an AI tool specifically designed for interviews and adopting a general AI model.
\textit{``If I had specifically created an AI model for interviews, I would consider it my fault if it had issues. But if it's just a general-purpose model, I think it's normal for it to make mistakes''} (P14).
\vspace{-0.4cm}
\subsection{Data and Privacy Protection}
Participants raised concerns about data security and privacy when introducing AGQs, particularly when AI can listen, record, or transcribe in real time. For instance, P12 emphasized incidental capture risk in public environments and noted they ``\textit{have no control over what should be captured}''.

They also viewed disclosure of AI involvement as ethically necessary, yet feared disclosure could jeopardize rapport and trust. For instance, P12 questioned whether interviewees would trust them or AI after being informed of the presence of AI. 

In addition to the dilemma between transparency and relational trust, the privacy issue is also highlighted. P12 noted that people with ``\textit{highly private experience}'' are willing to participate under conditions of anonymity and questioned that the presence of AI nearby might ``\textit{undermine trust.}''  In this context, privacy and trust were considered mutually reinforcing for stigmatized or highly sensitive experiences.

Beyond interpersonal concerns, some participants further highlighted the institutional and regulatory risks. Uploading interview data to third-party AI platforms was seen as violating institutional regulations and compliance requirements, particularly when sensitive information is involved (P14, P17).

\vspace{-0.2cm}
\section{Design and Governance Implications}

%\vspace{-0.2cm}
\begin{table*}[!h]
\centering
\small
\caption{Ethical risk categories and corresponding mitigation strategies for AI-assisted qualitative interviewing.}
\label{tab:risk_mitigation}
\vspace{-0.2cm}
\begin{tabular}{p{0.175\linewidth} p{0.195\linewidth} p{0.57\linewidth}}
\toprule
\textbf{Risk Category} & \textbf{Risk Description} & \textbf{Corresponding Mitigation Strategies} \\
\midrule

\textbf{Language Harm} &
Bias, offensive language, cultural insensitivity, misinterpretation; potential for emotional harm or misrepresentation. &
\textbf{Interaction Design Controls:}
Implement bias-aware prompts and filters using predefined ethical guidelines and real-time detection to mitigate biased language in AI-generated questions; design for active consent and transparency by clearly disclosing AI involvement and enabling explicit opt-out and granular control over data usage; incorporate empathy and rapport-building features, such as polite and empathetic language markers and conversational cues. \\

\addlinespace

\textbf{Respect \& Rapport} &
Erosion of trust, dehumanization, lack of empathy, failure to build meaningful connection; impersonal interaction. &
\textbf{Interaction Design Controls:}
Ensure inclusive interface and accessibility by optimizing platforms for diverse devices and abilities and providing accessible UI features; incorporate empathy and rapport-building features to support respectful and human-centered interaction. \\

\addlinespace

\textbf{Participation Inequality} &
Digital access barriers, unequal representation, power imbalances, exclusion of marginalized voices; unfair access. &
\textbf{Interaction Design Controls:}
Ensure inclusive interface and accessibility by supporting diverse devices, abilities, and contexts; reduce barriers that require specialized technical knowledge to engage with AI-assisted interviewing tools. \\

\addlinespace

\textbf{Responsibility Ambiguity} &
Unclear accountability for errors, diffused decision-making, lack of ownership for AI-generated outcomes; accountability gaps. &
\textbf{Human Oversight \& Accountability:}
Maintain human-in-the-loop review by mandating interviewer approval of all AI-generated content before delivery; establish clear accountability frameworks that define specific roles and responsibilities for human team members regarding system errors and ethical breaches; provide ethical training for interviewers on AI limitations, potential biases, and best practices; implement feedback mechanisms allowing participants and interviewers to report issues or provide feedback on AI interactions. \\

\addlinespace

\textbf{Privacy \& Compliance} &
Unauthorized data use, confidentiality breaches, non-compliance with regulations, surveillance concerns; data insecurity. &
\textbf{Data \& Governance Measures:}
Adopt privacy-by-design principles to minimize data collection, anonymize responses, and ensure secure data storage; enforce strict data retention and deletion policies with clear, time-bound protocols; conduct regular ethical audits and impact assessments to evaluate bias, fairness, and regulatory compliance; ensure compliance with relevant regulations (e.g., GDPR, CCPA) and ethical research guidelines. \\

\bottomrule

\end{tabular}
\end{table*}

We translate these findings into actionable guidance aligned with established human-subject ethics~\cite{united1978belmont} and contemporary AI risk management principles~\cite{10664609}. Rather than prescribing a single configuration for AI-assisted interviewing, we highlight key design and governance decision points that shape how ethical risks are managed in practice. Table.~\ref{tab:risk_mitigation} synthesizes the five risk themes identified in our analysis into a visual mitigation checklist, illustrating how ethical concerns can be addressed through coordinated interaction design choices, human oversight and accountability mechanisms, and data governance practices.

%\vspace{-0.2cm}
%\begin{table}[htbp]
%\small
%\centering
%\caption{From risks to design/governance controls.}
%\vspace{-0.2cm}
%\label{tab:implications}
%\begin{tabular}{@{}p{0.22\linewidth}p{0.72\linewidth}@{}}
%\toprule
%\textbf{Risk theme} & \textbf{Recommended controls} \\
%\midrule
%Language harm & Conservative generation policies; toxicity/discrimination filters; safe fallback when uncertainty is high; rapid ``repair'' suggestions (apology/reframe); human approve/hold/skip workflow. \\
%\addlinespace
%Respect \& rapport & Attention-light oversight (minimal gaze switching); interviewer-controlled timing and pacing; backstage-by-default in sensitive contexts; clear turn-taking to avoid rigid procedural feel. \\
%\addlinespace
%Participation inequality & Reduce ``prompt literacy'' requirements; plain-language options with rationale; accessibility and multilingual support; interviewer customization for community norms. \\
%\addlinespace
%Responsibility ambiguity & Provenance and audit logs (suggested/edited/asked); explicit role configuration; governance statements clarifying responsibility among researchers, institutions, and providers. \\
%\addlinespace
%Privacy \& compliance & Data minimization; institution-approved processing; retention/redaction controls; clear consent and disclosure with opt-out; avoid incidental capture; compliance-aligned deployment settings. \\
%\bottomrule
%\end{tabular}
%\end{table}

\vspace{-0.2cm}
\subsection{Practical Checklist for AI-Assisted Interviewing}
To support practical adoption, we further distill these implications into a lightweight checklist that reflects how ethical considerations unfold across the temporal structure of an interview. This checklist is intended as a heuristic rather than a rigid protocol, supporting situated judgment by researchers and tool builders.

\textbf{Before the interview}, researchers and tool builders should configure conservative language guardrails, define clear disclosure practices and opt-out options, and ensure compliance for data storage and transcription. Establishing explicit human-override protocols in advance can help interviewers respond quickly to uncertain or potentially harmful AI-generated suggestions without disrupting the interaction. 

\textbf{During the interview}, AI suggestions should be positioned backstage unless their inclusion is explicitly appropriate. Interface designs should minimize attention-diverting interactions and support lightweight approve/hold/skip workflows that allow interviewers to retain control over pacing and content. Throughout the interaction, maintaining rapport and respecting interpersonal boundaries should take precedence over maximizing AI involvement. 

\textbf{After the interview}, teams should maintain audit trails documenting how AI suggestions were used, review flagged or uncertain outputs, and explicitly document how responsibility is allocated among human and system actors. Incorporating participant feedback on disclosure, trust, and comfort can further inform iterative refinement of AI-assisted interviewing practices.

%\begin{figure}[htbp]
%  \centering
%  \includegraphics[width=0.98\linewidth]{graph/Risk to Mitigation Checklist.png}
%  \vspace{-0.2cm}
%  \caption{Mitigation checklist. Five ethical risk categories are linked to corresponding mitigation strategies across interaction design, human oversight, and data governance. The framework illustrates how ethical concerns can be translated into actionable design and institutional safeguards.}
%  \label{fig:mitigation}
%\end{figure}
\vspace{-0.2cm}
\section{Limitations and Future Work}
This LLM-in-the-loop WoZ study approximates real deployment while preserving experimental control and human oversight. Future work should evaluate these ethical risks in longitudinal use, higher-stakes domains, and cross-cultural settings where norms of respect, disclosure, and trust vary. Additional work is needed to develop governance models that make responsibility attribution actionable across researchers, institutions, and AI providers. Notably, the prompt used in this study was intentionally simplified, asking GPT-4o only to generate follow-up questions based on the current conversation. Future systems and interaction models may benefit from more comprehensive design.
\vspace{-0.2cm}
\section{Conclusion}
Integrating AI-generated follow-up questions into semi-structured interviews introduces ethical and social responsibility challenges beyond question quality. Our new findings highlight risks around harmful language, respect and rapport, participation inequality, responsibility attribution, and privacy/compliance. We provide a risk-to-controls mapping that can guide safer, more respectful, and more accountable AI-assisted interviewing.

%\subsection{}
\vspace{-0.2cm}
\begin{acks}
We thank the anonymous reviewers and all participants for their time, insights, and thoughtful reflections. This work was approved by the Institutional Review Board (IRB) at the first author's institution (STUDY00023010). We also acknowledge the use of generative AI tools in this work. LLMs were used to support language refinement, such as grammar and spelling checks, during manuscript preparation. The authors take full responsibility for the design, interpretation, and presentation of all content, as well as for the appropriate and ethical use of AI in this work. This work was supported by the Center for Socially Responsible Artificial Intelligence (CSRAI) of the Pennsylvania State University.
\end{acks}
%%
%% The acknowledgments section is defined using the "acks" environment
%% (and NOT an unnumbered section). This ensures the proper
%% identification of the section in the article metadata, and the
%% consistent spelling of the heading.
%\begin{acks}

%\end{acks}

%%
%% The next two lines define the bibliography style to be used, and
%% the bibliography file.
%\FloatBarrier
\vspace{-0.2cm}
\balance
\bibliographystyle{ACM-Reference-Format}
\bibliography{references,references_Iris,main}

%%
%% If your work has an appendix, this is the place to put it.
\appendix

\end{document}